\documentclass[conference]{IEEEtran}
\IEEEoverridecommandlockouts
\usepackage[numbers,compress]{natbib}
\usepackage{url}
\usepackage{amsmath,amssymb,amsfonts}
\usepackage[capitalize]{cleveref}
\usepackage{algorithmic}
\usepackage{graphicx}
\usepackage{textcomp}
\usepackage{xcolor}
\usepackage{siunitx}
\usepackage[shortcuts,acronym,automake=false]{glossaries}
\def\BibTeX{{\rm B\kern-.05em{\sc i\kern-.025em b}\kern-.08em
    T\kern-.1667em\lower.7ex\hbox{E}\kern-.125emX}}

\begin{document}
\newacronym{isac}{ISAC}{integrated sensing and communication}
\newacronym{tx}{TX}{transmission}
\newacronym{rx}{RX}{reception}
\newacronym{if}{IF}{intermediate frequency}
\newacronym{rf}{RF}{radio frequency}
\newacronym{lo}{LO}{local oscillator}
\newacronym{pri}{PRI}{pulse repetition interval}
\newacronym{cpi}{CPI}{coherent processing interval}

\title{Demonstration of a Real-Time Testbed for D-Band Integrated Sensing and Communication
\thanks{The authors of this work acknowledge the financial support by the Federal Ministry of Education and Research of Germany (BMBF) in the programme ``Souverän. Digital. Vernetzt.'' joint project 6G-RIC (grant numbers: 16KISK020K, 16KISK030).}
}

\author{\IEEEauthorblockN{Sven Wittig\IEEEauthorrefmark{1},
Rodrigo Hernang\'{o}mez\IEEEauthorrefmark{1}, Karen Vardanyan\IEEEauthorrefmark{3}, Ramez Askar\IEEEauthorrefmark{1}, Amr Haj-Omar\IEEEauthorrefmark{3},
Michael Peter\IEEEauthorrefmark{1}, \\S{\l}awomir Sta\'{n}czak\IEEEauthorrefmark{1}\IEEEauthorrefmark{2}}

\IEEEauthorblockA{\IEEEauthorrefmark{1}Fraunhofer Heinrich Hertz Institute, Berlin, Germany, \{firstname.lastname\}@hhi.fraunhofer.de}
\IEEEauthorblockA{\IEEEauthorrefmark{2}
Technische Universit\"{a}t Berlin, Berlin, Germany, \{firstname.lastname\}@tu-berlin.de}
\IEEEauthorblockA{\IEEEauthorrefmark{3}NI, Austin, TX, USA, \{firstname.lastname\}@ni.com}
}

\maketitle

\begin{abstract}
The D-band, spanning 110 GHz to 170 GHz, has emerged as a relevant frequency range for future mobile communications and radar sensing applications, particularly in the context of 6G technologies. This demonstration presents a high-bandwidth, real-time \gls{isac} platform operating in the upper D-band at \SI{160}{\giga\hertz}. The platform comprises a software-defined \acrlong{if} transceiver and a D-band \acrlong{rf} module. Its flexible software design allows for rapid integration of signal processing algorithms. Highlighting its potential for advanced research in ISAC systems, the platforms's efficacy is showcased through a live demonstration with an algorithm implementing human target tracking and activity classification.
\end{abstract}


\glsresetall

\section{Introduction}

The D-band, ranging from \SIrange{110}{170}{\giga\hertz}, has been identified as a promising radio frequency band for new-generation mobile communication standards in 6G and beyond~\cite{eichler2022fundamentals}.
The availability of large bandwidths in this band does not only allow higher data rates for wireless communication, but it can also enable high-resolution radar sensing.
Thus, D-band frequencies constitute an interesting subject of study from the viewpoint of \gls{isac}, which has emerged as a widely considered research topic towards new mobile standards~\cite{liu_integrated_2022}.

In this demonstration, we show a high bandwidth, real-time capable \gls{isac} research platform operating in the upper D-band. To show its capabilities, we implement a basic human target tracking and activity classification algorithm.

\section{D-Band ISAC Testbed System Design}

\subsection{Setup \& Data Acquisition}

\begin{figure}
    \centering
    \includegraphics[width=\columnwidth]{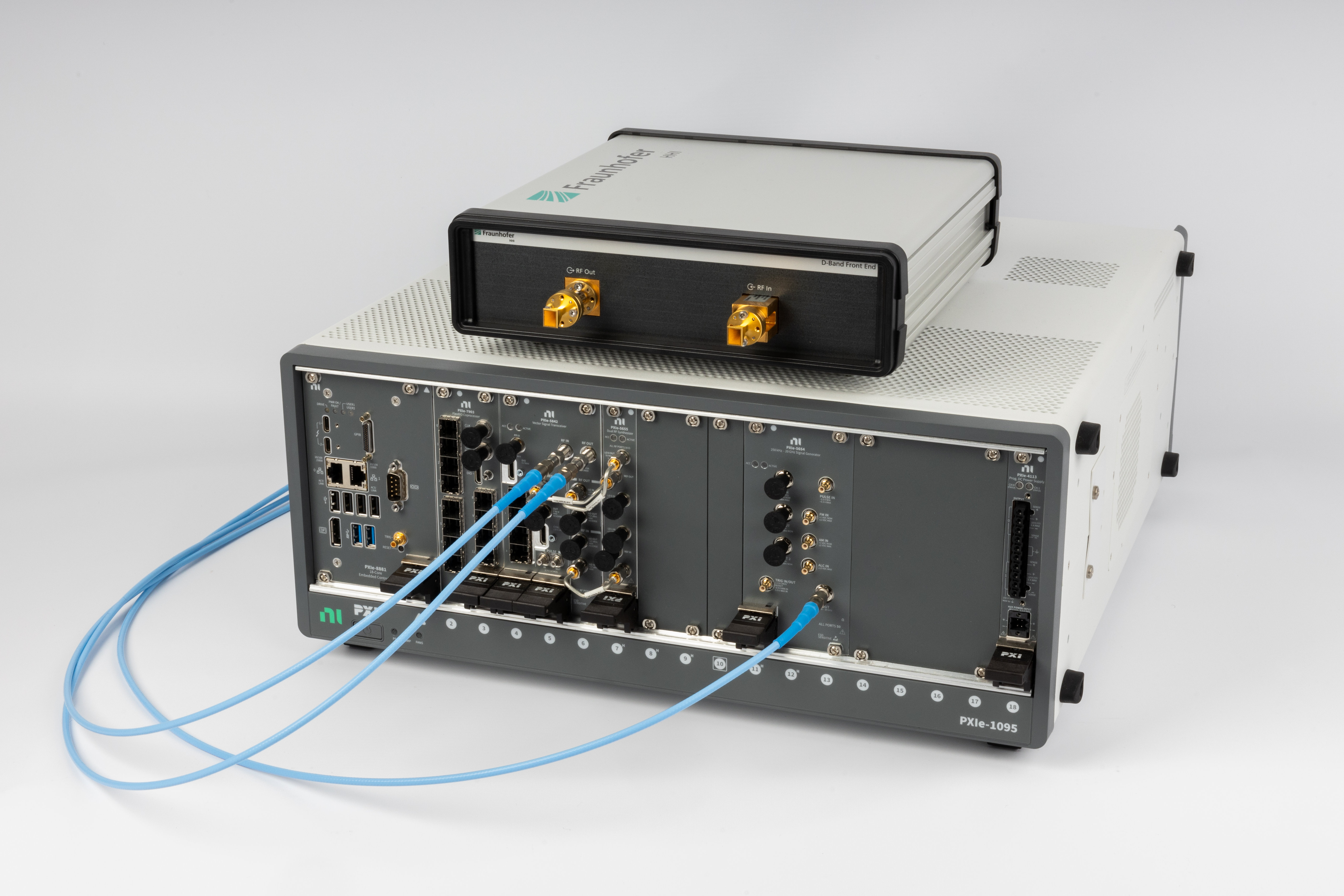}
    \caption{The hardware of the D-band \gls{isac} testbed.}
    \label{fig:hardware}
    \vspace{-1em}
\end{figure}

\begin{figure*}
    \vspace{0.04in}
    \centering
    \includegraphics[width=1.75\columnwidth]{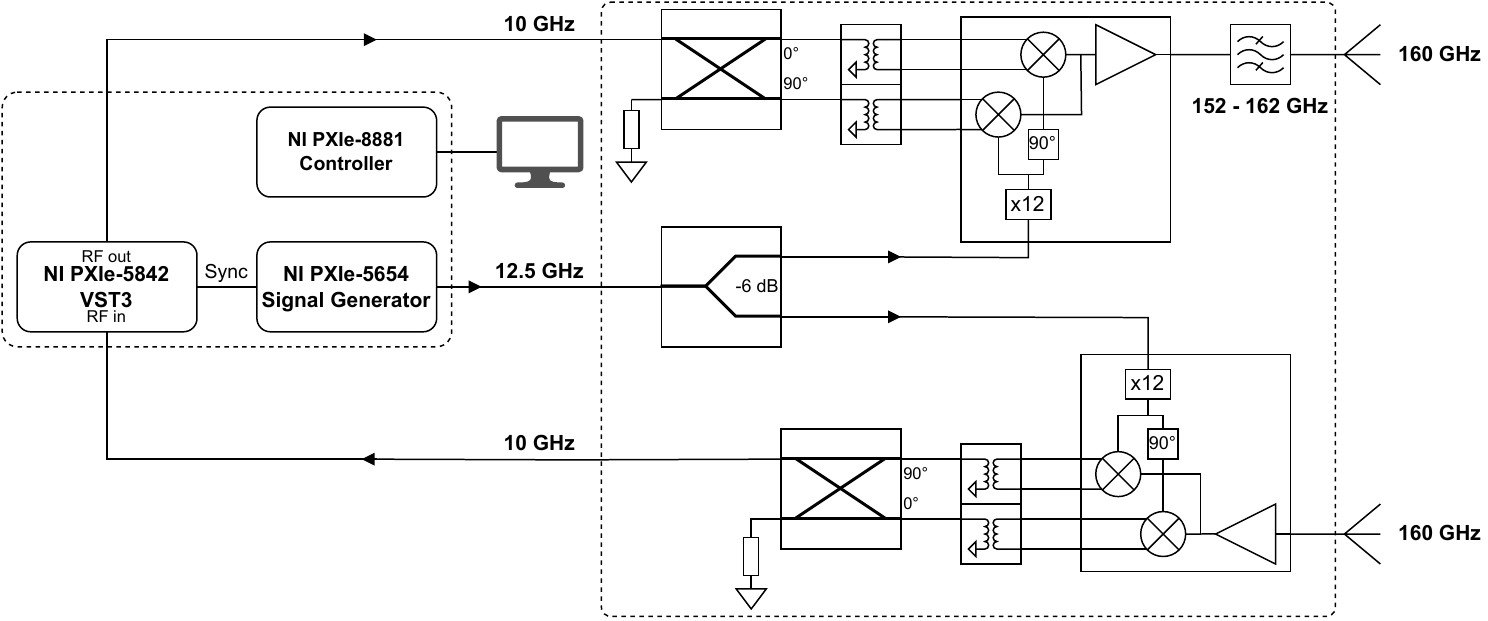}
    \caption{Block diagram of the testbed hardware, including the PXIe based IF transceiver and \gls{lo} generation (left), and  the \gls{rf} up-down-converter (right).}
    \label{fig:blockdiagram}
\end{figure*}

The demonstrated D-band \gls{isac} testbed as depicted in \cref{fig:hardware} and \cref{fig:blockdiagram} consists of a software-defined \gls{if} transceiver and an integrated single-sideband up-down-converter module.

Signal transmission and reception is realized using an NI PXIe-5842 Vector Signal Transceiver at an \gls{if} of \SI{10}{\giga\hertz} with \SI{4}{\giga\hertz} digital baseband bandwidth. Signal processing is implemented in software in a PXIe-8881 embedded controller. The controller runs an NI LabVIEW application that acquires and displays range-Doppler profiles in real time. It simultaneously passes the raw range profiles on to a Python environment for further processing, in this demonstration detection, tracking and classification. The backend Python code can be modified on-the-fly, making the testbed easily accessible to signal processing researchers.

The D-band \gls{rf} module is assembled from discrete waveguide and connectorized components, using two IQ mixers with \gls{if}-side quadrature hybrids to realize single-sideband up- and downconversion. The \gls{lo} signal at \SI{12.5}{\giga\hertz} is supplied by an NI PXIe-5654 signal generator and multiplied internally to \SI{150}{\giga\hertz}.

While the testbed supports arbitrary waveforms, the demonstration uses  $8192$-sample Zadoff-Chu sequences, which are also employed for synchronization and random access in mobile communication, to acquire range profiles by digital correlation. With a \acrlong{pri} of 100 sequence periods (or around \SI{0.2}{\milli\second}) and a \acrlong{cpi} of 512 pulses, this configuration gives a sample-based range resolution of \SI{3.75}{\centi\meter} and a velocity resolution of \SI{0.9}{\centi\meter\per\second} at a maximum velocity of \SI{2.29}{\meter\per\second}. The useful range is limited by link power budget, not ambiguity.

\subsection{Data Processing}

\begin{figure}
    \centering
    \includegraphics[width=\linewidth]{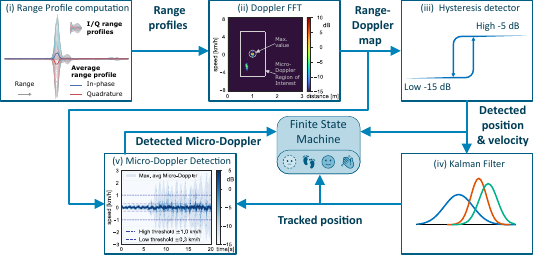}
    \caption{Block diagram of the detection, tracking and classification algorithm.}
    \label{fig:algo}
    \vspace{-1em}
\end{figure}

The tracking and classification algorithm combines 5 processing stages with a finite state machine, as illustrated in \cref{fig:algo}. After range profile acquisition (\cref{fig:algo}(i)), a Fourier transform is computed across 512 range profiles with a Hann window to produce a range-Doppler map (\cref{fig:algo}(ii)). Prior to that, the zero-Doppler clutter is estimated as the average across range profiles (cf. \cref{fig:algo}(i)) and subtracted from them.

In order to detect a single moving target, we extract the maximum value within the range-Doppler scope shown in \cref{fig:algo}(ii) and feed it into a hysteresis detector with fixed thresholds (\cref{fig:algo}(iii)). Upon detection, the target's position and velocity are fed into a Kalman filter (\cref{fig:algo}(iv)). Micro-Doppler features \cite{chen2011microdoppler} are then obtained based on the range-Doppler map and Kalman filter's output (\cref{fig:algo}(v)). In particular, we extract the maximum absolute velocity for which the return power exceeds
\SI{-15}{\deci\bel} within a region of interest around the tracked position and velocity (cf. \cref{fig:algo}(ii)). The micro-Doppler velocity is finally computed as a moving average of length \SI{0.5}{\second} over the maximum absolute velocity.

The finite state machine in \cref{fig:algo} uses the outputs from \cref{fig:algo}(iii)-(v) to determine the target state as either \emph{absent}, \emph{standing}, \emph{walking}, or \emph{waving}. More specifically, the state remains \emph{absent} unless a target is detected in \cref{fig:algo}(iii). In that case, the state changes to \emph{walking} if the tracked position drifts more than \SI{10}{\centi\meter} within \SI{0.5}{\second}. Otherwise, the micro-Doppler velocity is compared against two hysteresis thresholds (cf. \cref{fig:algo}(v)) to decide whether the target is \emph{standing} still or performing \emph{hand-waving}-like gestures.

\section{Live Demonstration: Target Tracking \& Classification}

\begin{figure}
    \centering
    \includegraphics[width=0.925\columnwidth]{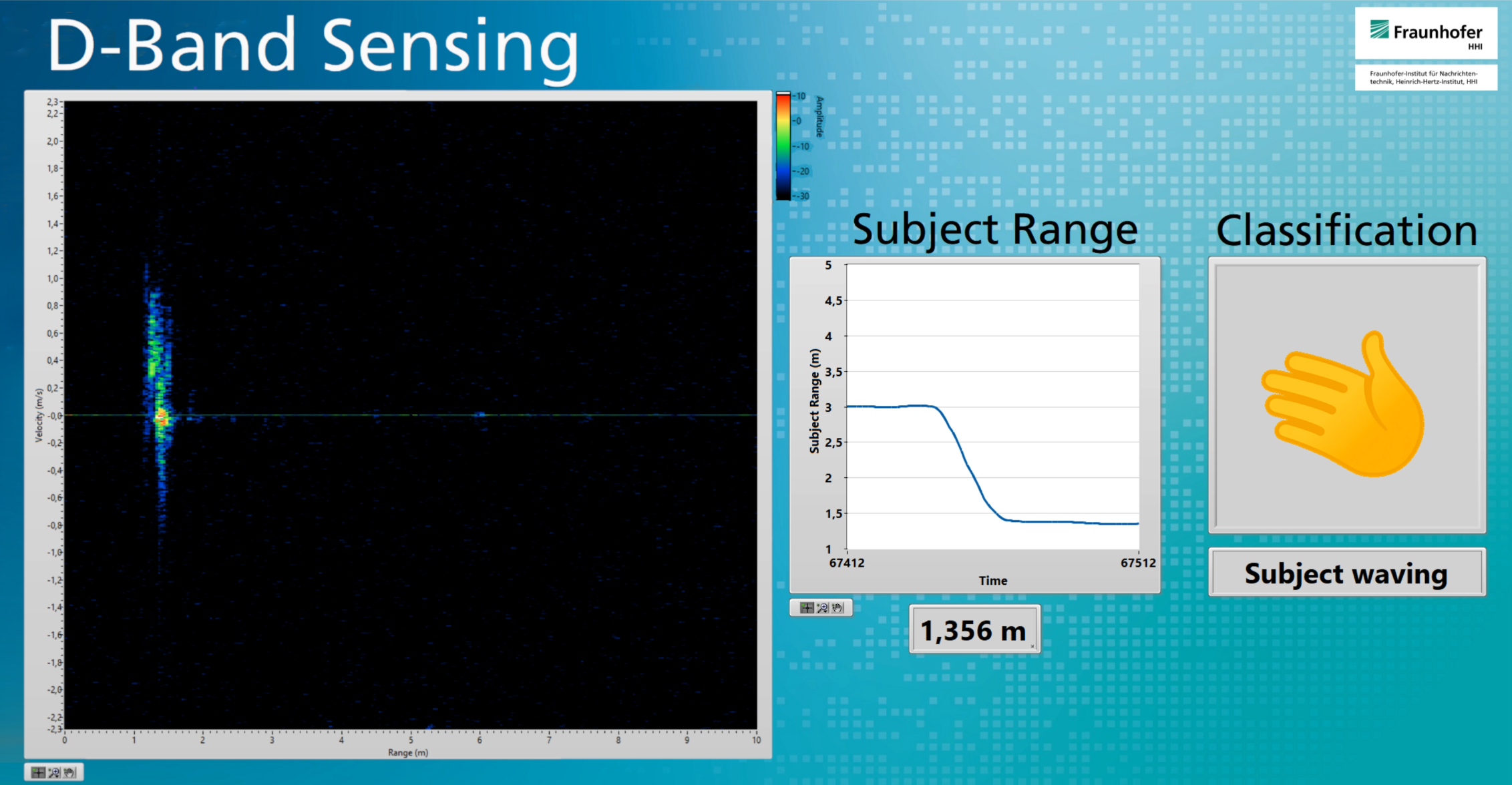}
    \caption{Graphical interface of the demonstration system.}
    \label{fig:gui}
    \vspace{-1em}
\end{figure}

The demonstrations's graphical interface, depicted in \cref{fig:gui}, shows the range-Doppler map together with the tracked position and classification output in real time. It allows direct interaction with the demonstration in a live environment. \footnote{Video available at \url{https://s.fhg.de/hhi-dband-sensing}}

\bibliographystyle{IEEEtranN}
\bibliography{IEEEabrv,references}

\end{document}